\documentclass[a4paper,10pt,twocolumn]{article}

\usepackage[utf8]{inputenc}
\usepackage[T1]{fontenc}
\usepackage[english]{babel}
\usepackage{xcolor}
\usepackage{ifthen}
\usepackage{calc}

\usepackage{times}      
\usepackage{ifpdf}

\usepackage{amsmath,amsfonts,amssymb}
\usepackage{graphicx} 
\usepackage{xcolor}
\usepackage{booktabs}
\usepackage{siunitx}
\usepackage{stfloats}

\usepackage{authblk}
\usepackage[left=2cm,%
                right=2cm,%
                top=2.25cm,%
                bottom=2.25cm,%
                headheight=12pt,%
                letterpaper]{geometry}%
                
\usepackage[labelfont={bf,sf},%
                labelsep=space,%
                textfont=sf,%
                justification=justified,font=small]{caption}

\usepackage[colorlinks=true, allcolors=blue]{hyperref}

\usepackage[superscript,biblabel,nomove]{cite}
\usepackage{fancyhdr}  
\usepackage{lastpage}  
\usepackage[explicit]{titlesec}
\titleformat{\section}
  {\color{color1}\large\sffamily\bfseries}
  {\thesection}
  {0.5em}
  {#1}
  []
\titleformat{name=\section,numberless}
  {\color{color1}\large\sffamily\bfseries}
  {}
  {0em}
  {#1}
  []  
\titleformat{\subsection}
  {\sffamily\bfseries}
  {\thesubsection}
  {0.5em}
  {#1}
  []
\titleformat{\subsubsection}
  {\sffamily\small\bfseries\itshape}
  {\thesubsubsection}
  {0.5em}
  {#1}
  []    
\titleformat{\paragraph}[runin]
  {\sffamily\small\bfseries}
  {}
  {0em}
  {#1} 
\titlespacing*{\section}{0pc}{3ex \@plus4pt \@minus3pt}{5pt}
\titlespacing*{\subsection}{0pc}{2.5ex \@plus3pt \@minus2pt}{0pt}
\titlespacing*{\subsubsection}{0pc}{2ex \@plus2.5pt \@minus1.5pt}{0pt}
\titlespacing*{\paragraph}{0pc}{1.5ex \@plus2pt \@minus1pt}{10pt}

\usepackage{titletoc}
\contentsmargin{0cm}
\titlecontents{section}[\tocsep]
  {\addvspace{4pt}\small\bfseries\sffamily}
  {\contentslabel[\thecontentslabel]{\tocsep}}
  {}
  {\hfill\thecontentspage}
  []
\titlecontents{subsection}[\tocsep]
  {\addvspace{2pt}\small\sffamily}
  {\contentslabel[\thecontentslabel]{\tocsep}}
  {}
  {\ \titlerule*[.5pc]{.}\ \thecontentspage}
  []
\titlecontents*{subsubsection}[\tocsep]
  {\footnotesize\sffamily}
  {}
  {}
  {}
  [\ \textbullet\ ]  
  
\usepackage{enumitem}

\renewcommand{\@biblabel}[1]{\bfseries\color{color1}#1.}

\newcommand{\keywords}[1]{\def\@keywords{#1}}

\def\xabstract{abstract}
\long\def\abstract#1\end#2{\def\two{#2}\ifx\two\xabstract 
\long\gdef\theabstract{\ignorespaces#1}
\def\go{\end{abstract}}\else
\typeout{^^J^^J PLEASE DO NOT USE ANY \string\begin\space \string\end^^J
COMMANDS WITHIN ABSTRACT^^J^^J}#1\end{#2}
\gdef\theabstract{\vskip12pt BADLY FORMED ABSTRACT: PLEASE DO
NOT USE {\tt\string\begin...\string\end} COMMANDS WITHIN
THE ABSTRACT\vskip12pt}\let\go\relax\fi
\go}

\renewcommand{\@maketitle}{%
{%
\thispagestyle{empty}%
\vskip-36pt%
{\raggedright\sffamily\bfseries\fontsize{18}{22}\selectfont \@title\par}%
\vskip10pt
{\raggedright\sffamily\fontsize{12}{16}\selectfont  \@author\par}
\vskip10pt
{%
\noindent
\colorbox{color2}{%
\parbox{\dimexpr\linewidth-2\fboxsep\relax}{%
\sffamily\small\textbf\\\theabstract
}%
}%
}%
\vskip25pt%
}%
}%
\definecolor{color1}{RGB}{0,0,0} 
\definecolor{color2}{gray}{1} 

\newlength{\tocsep} 
\usepackage{xcolor}

\let\oldbibliography\thebibliography
\renewcommand{\thebibliography}[1]{%
\addcontentsline{toc}{section}{\hspace*{-\tocsep}\refname}%
\oldbibliography{#1}%
\setlength\itemsep{0pt}%
}

\usepackage{nccmath}
\usepackage{ulem} 
\usepackage{verbatim}

\title{Impact of spectral effects on photovoltaic \\ energy production: A case study in the United States
}

\author[1,*]{Jos\'{e}~M.~Ripalda}
\affil[1]{Instituto de Micro y Nanotecnología - CSIC, Isaac Newton, 8, E-28760, Tres Cantos, Madrid, Spain}
\affil[*]{j.ripalda@csic.es}
\author[2]{Daniel~Chemisana}
\affil[2]{Applied Physics Section of the Environmental Sci. Dept, Universitat de Lleida, Jaume II 69, 25001, Lleida, Spain}
\author[1]{Jos\'{e} M. Llorens}
\author[3]{Iv\'{a}n~Garc\'{i}a}
\affil[3]{Instituto de Energía Solar, Universidad Politécnica de Madrid, Avda. Complutense 30, 28040, Madrid, Spain}

\begin{abstract}
\textbf{The time averaged efficiency of photovoltaic modules in the field is generally lower than the efficiency under standard testing conditions due to the combined effects of temperature and spectral variability, affecting the bankability of power plant projects. We report spectral correction factors ranging from -2\% to 1.3\% of the produced energy for silicon modules depending on location and collector geometry. We find that spectral effects favor trackers if silicon modules are used, but favor a fixed tilt instead if perovskites or CdTe are used. In high irradiance locations, the energy yield advantage of silicon based trackers is underestimated by 0.4\% if spectral sensitivity effects are neglected. As the photovoltaic market grows to a multi-terawatt size, these seemingly small effects are expected to have an economic impact equivalent to tens of billions of dollars in the next few decades, far out-weighting the cost of the required research effort.}
\end{abstract}

\begin{document}

\flushbottom
\maketitle
\thispagestyle{empty}

Due to the rapid cost reduction of photovoltaics (PV), recent forecasts are predicting that several tens of terawatts of PV capacity will be deployed before 2050\cite{science_terawatt_2019}. This represents an investment of several tens of trillions of dollars. As a consequence there is a large economic drive to optimize the choice of location and technology for new PV systems. Key aspects to take into account are the geographical and temporal variations of the spectral irradiance, and meteorological parameters such as ambient temperature and wind speed. Changes in the spectral irradiance are mostly driven by the position of the sun and atmospheric conditions\cite{kurtz_difference_1991, faine_influence_1991, kurtz_projected_1997, chan_impact_2014, ekins-daukes_brighten_2019, sweerts_estimation_2019}, but also by the orientation of solar panels as defined by the plane of array (POA). Accurately accounting for these effects requires detailed radiative transfer models including multiple reflection, scattering, and absorption events in the atmosphere including both cloudy and clear-sky conditions\cite{xie_fast_2018, xie_fast_2019}. Data from these radiative transfer models has only recently become widely available through the National Solar Radiation DataBase  (NSRDB) web service\cite{xie_fast_2018, xie_fast_2019, NSRDB_2018}. Here we use these spectral and meteorological datasets to obtain the PV efficiency and energy production as a function of location for a wide range of PV technologies. Previous studies have found that spectral variability has a significant effect on the energy production, but have not compared tracking with fixed tilt collection geometries\cite{ripalda_solar_2018, vossier_is_2017, garcia_spectral_2018, dirnberger, edu, pvmaps, kinsey, huld_estimating_2015, lindsay_errors_2020, peters_global_2018, warmann_predicting_2019}. Here we include the effects of wind, ambient temperature, and irradiance on solar cell temperatures in addition to spectral variability including the effects of clouds. Our results exemplify that consideration of the combined effects of spectral and temperature variations will allow to fine tune the optimal location, module technology, and collection geometry for each PV project, with an economic benefit far out-weighting the cost of the required research effort. Most importantly, we provide spectral correction factors, for each location and PV technology, that can be used to improve the accuracy of conventional energy production forecasts.

In the first section of this work we are concerned with the implications of thermal and spectral variability for mainstream PV technology based on fixed tilt silicon modules. Thin film technologies such as CdTe and perovskites are also discussed. We then examine the implications of our study for the energy production of tracking systems. In the next section we also consider multijunctions under global spectra. We then quantify the band gap adjustments required at specific locations to maximize the produced energy. To conclude we discuss the uncertainty in our results as a function of the number of spectra used per location.

\begin{figure}[th!]
\centering
    \includegraphics[width=0.47\textwidth]{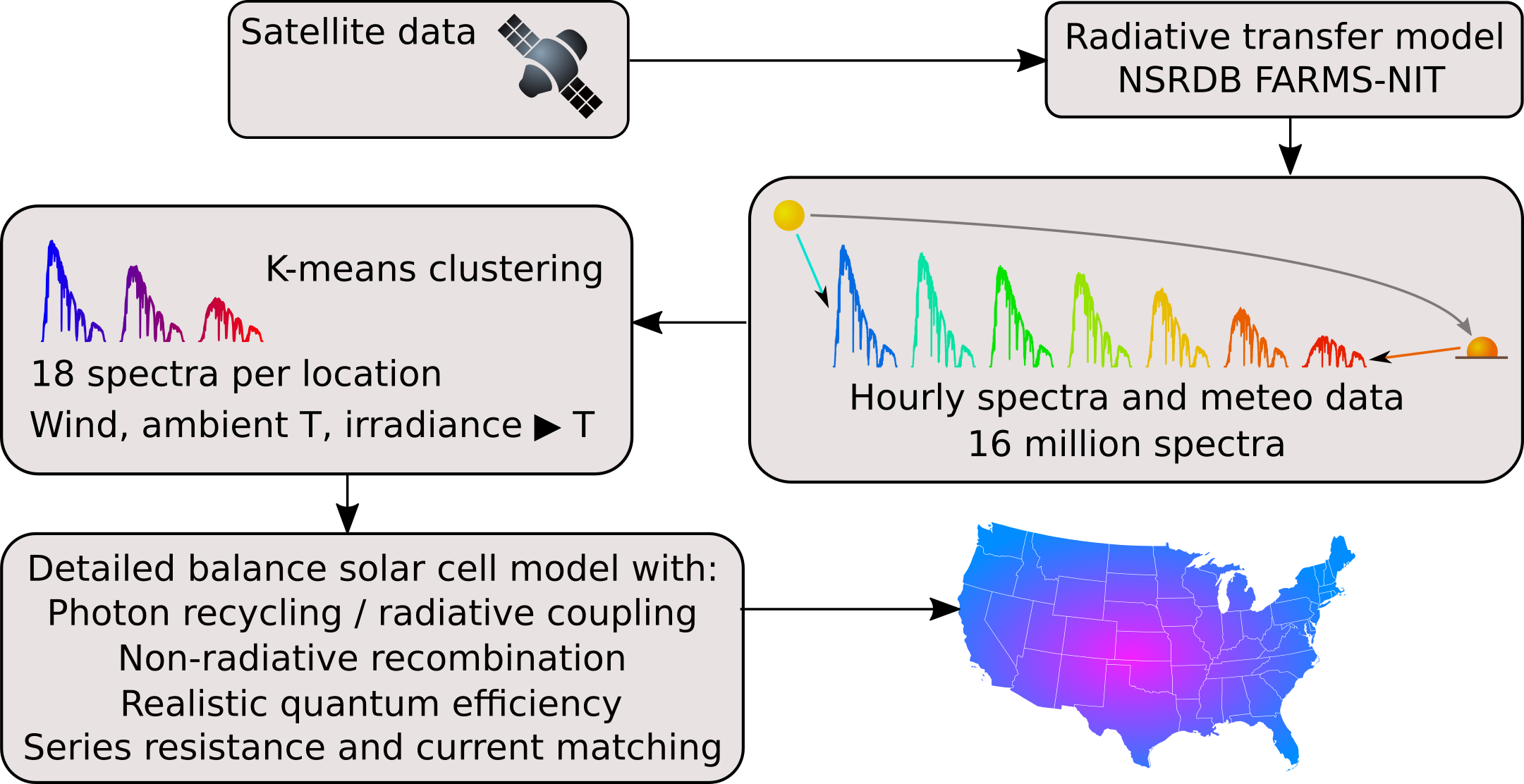}
    \caption{Flow chart of the methodology to calculate the yearly energy yield and time averaged PV efficiency as a function of location. Further details are given in the Methods section.}
\label{fig:flow}
\end{figure}

A flow chart summarizing our methodology is shown in Fig. \ref{fig:flow}. We have included in our calculations the most relevant effects as detailed in the methods section and Ref. \cite{ripalda_solar_2018}. We use the Sandia PV Array Performance Model for solar cell temperatures using ambient temperature and wind speed data\cite{sandia_2004}. In single junctions, the most pronounced effect of temperature is a reduction in the voltage due to a higher recombination current\cite{dupre_physics_2015, green_thermal_2017}. An appropriate model for solar cell temperatures is also required because of the Varshni shift of the bandgaps with temperature.

\section*{Silicon}
\begin{figure}[th!]
\centering
    \includegraphics[width=0.47\textwidth]{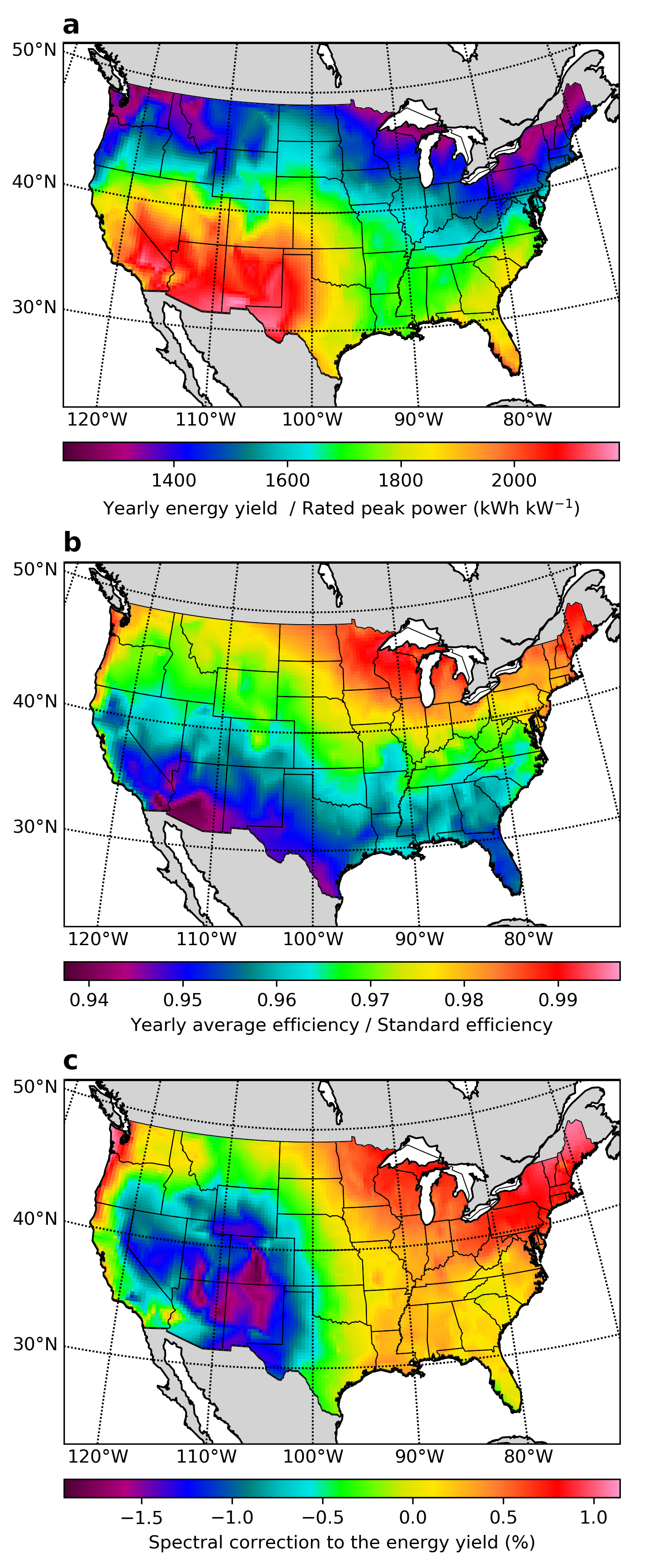}
    \caption{Silicon single junctions at a fixed optimal tilt angle. \textbf{a}, Yearly energy production relative to the rated peak power of the PV system. \textbf{b}, Yearly averaged efficiency relative to the standard efficiency. \textbf{c}, Spectral correction factors to be used when calculating the efficiency or the energy production assuming the standard spectrum.}
\label{fig:si}
\end{figure}

A typical PV system uses silicon modules fixed at a tilt angle roughly matching the latitude and oriented towards the south (or to the north if in the southern hemisphere). We adjust the tilt angle for each location as a function of latitude according to the prescription given by Jacobson and Jadhav\cite{jacobson_world_2018}. We have calculated the maximum realistically achievable yearly energy production (Fig. \ref{fig:si}a) and yearly averaged energy efficiency (Fig. \ref{fig:si}b) for such systems when considering spectral and temperature variability effects. The trends in the energy production map are opposite to those in the efficiency map due to the effect of higher solar cell temperatures in high irradiance locations. The lower efficiency in the south is mostly due to the effect of temperature on the recombination current, and consequently on the voltage, but high temperatures also further shift the silicon band gap away from the optimal value for a single junction (1.35 eV according to Ref. \cite{ripalda_solar_2018}). Temperature effects slightly reduce the economic advantage of deployment in high irradiance locations.

To quantify spectral sensitivity effects we compute spectral correction factors $f_{s}$ as the ratio of the yearly energy yield $E_\text{y}$ obtained with the NSRDB POA spectral irradiance $s(\lambda)$ and the yearly energy yield obtained assuming the standard ASTM G-173-03 spectrum $s_0(\lambda)$ scaled to match the NSRDB POA irradiance $G = \int_{0}^{\infty} s(\lambda) \, d \lambda$ as

\begin{equation}
f_{s} = \frac{E_\text{y}(s(\lambda))}{E_\text{y}(s_{0}(\lambda) \, G/G_{0})},
\end{equation}
where $G_0$ is the integrated irradiance of the standard spectrum. To clearly separate spectral effects, the spectra are the only difference between these two energy yield calculations. Because the standard solar spectrum is often assumed to forecast the expected energy yield of new PV power plants, these spectral correction factors can be used to correct such forecasts. But these spectral correction factors can also be used to illustrate the relative importance of spectral sensitivity effects for each location and type of PV system. We present in Fig. \ref{fig:si}c the resulting spectral correction percentage as $(f_{s} - 1)$. Neglecting spectral effects thus leads to overestimating the energy yield in some of the locations with the highest production potential by nearly 2\% (high altitude locations in Colorado and New Mexico), while slightly underestimating it in others (the Sonoran Desert at the border between California and Arizona).

Spectral variability effects in single junctions are due to the absorption threshold of the semiconductor. These effects show a clear correlation between topographic altitude and efficiency losses in Fig. \ref{fig:si}c. By comparing the spectra at low altitude locations with the spectra from locations at high altitude, we observe that the efficiency is highest at low altitude due to higher infrared losses caused mostly by the water content of the atmosphere. Because these losses occur at energies below the band gap, they have the effect of an apparent efficiency increase that is not necessarily accompanied by an energy yield increase.

\section*{CdTe and perovskites}

\begin{figure}[bh!]
\centering
    \includegraphics[width=0.47\textwidth]{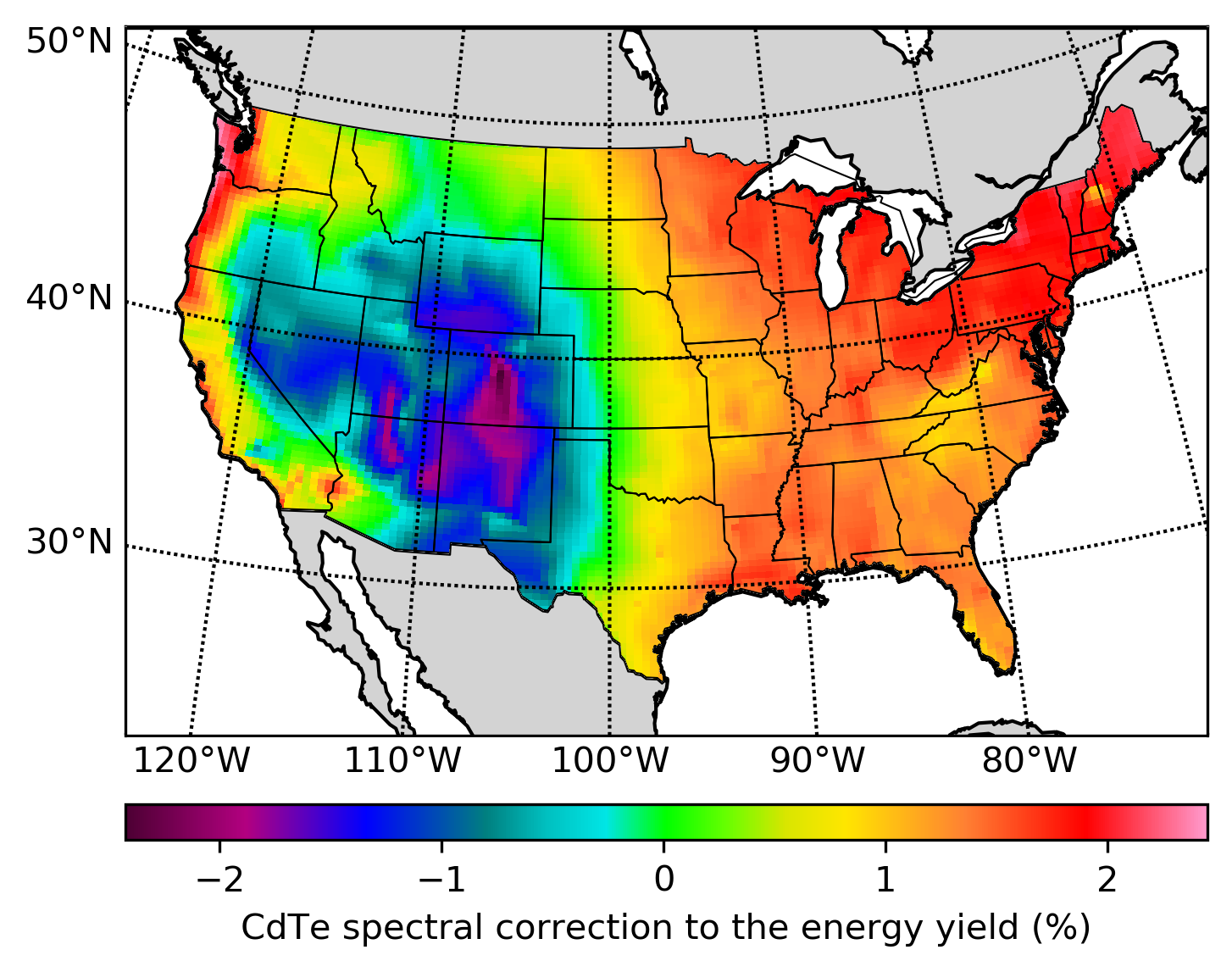}
    \caption{CdTe yearly spectral correction factor to the energy yield and the efficiency.}
\label{fig:cdte}
\end{figure}

As a consequence of the rapid drop in price of silicon based PV modules with higher efficiencies, the market share of thin film PV technologies based on CdTe and CuGaInSe$_2$ has declined slightly in recent years, but thin film technologies are likely to maintain a foothold in certain markets, applications or geographical regions. Here we center our attention on the case of CdTe, as its higher band gap (1.45 eV for CdTe vs 1.12 eV for silicon) might make it advantageous in locations with lower infrared irradiance or higher temperatures. The spectral correction factors in Fig. \ref{fig:cdte} do indeed show a wide geographical variation range.

In comparison with silicon, the performance of CdTe modules is more dependent on spectral effects due to its higher bandgap. As recently suggested by Peters \textit{et al.}\cite{peters_global_2018}, this is related to infrared losses in the atmosphere caused by water, decreasing the POA irradiance without decreasing the energy yield as these changes occur at energies below the CdTe bandgap. As a consequence the efficiency of CdTe and perovskite single junctions increases with increasing precipitable water in the atmosphere. 

\begin{figure}[bh!]
\centering
    \includegraphics[width=0.47\textwidth]{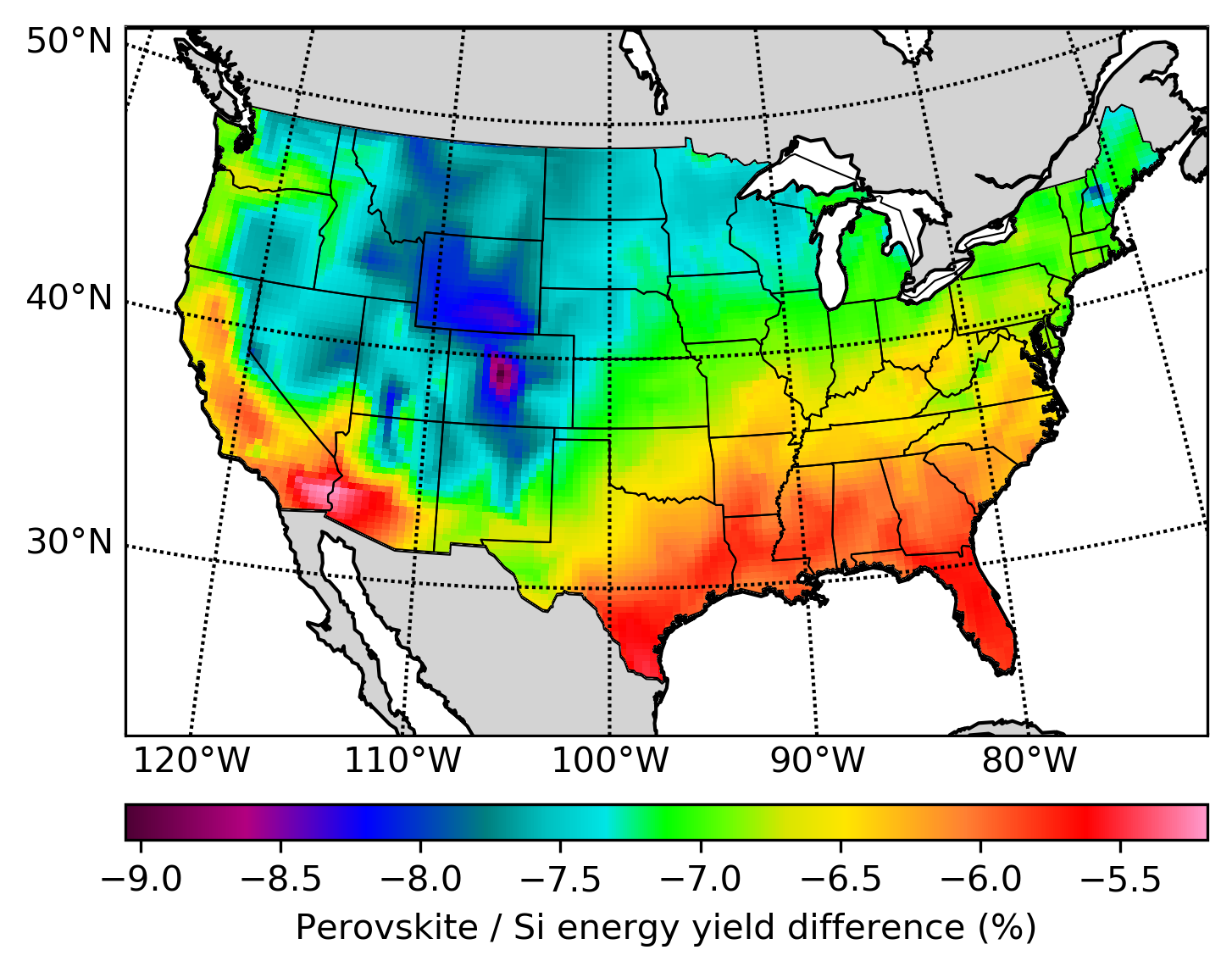}
    \caption{Relative energy yield difference between perovskite single junctions and silicon.}
\label{fig:p}
\end{figure}

Due to their similar bandgap, the geographical distributions of the energy yield, the efficiency, and the spectral correction factor, are very nearly the same for CdTe and perovskites, but with lower non-radiative recombination losses favoring perovskites over CdTe as we have optimistically assumed the performance parameters of record perovskite solar cells before degradation\cite{jung_efficient_2019}. Further details are given in the methods section. To compare the yearly energy yield of perovskite modules $E_\text{y}^\text{P}$ with the energy yield of silicon $E_\text{y}^\text{Si}$ in a fixed optimal tilt geometry, we plot in Fig. \ref{fig:p} the relative energy yield difference between perovskites and silicon as $E_\text{y}^\text{P} / E_\text{y}^\text{Si} - 1$. The POA irradiance is the same in both cases because the collection geometry is the same, and consequently the efficiency ratio is the same as the energy yield ratio (this will not be the case when studying the effect of tracking). We have assumed here the bandgap of the current record solar cell (1.5 eV)\cite{jung_efficient_2019}. The relevant result in Fig. \ref{fig:p} is the relative difference between locations, and not the absolute result at each location, as the actual performance of perovskite modules in the field is still largely unknown and we have not included time dependent degradation effects\cite{tress_performance_2019}.

\section*{Tracking}

\begin{figure}[h]
\centering
    \includegraphics[width=0.47\textwidth]{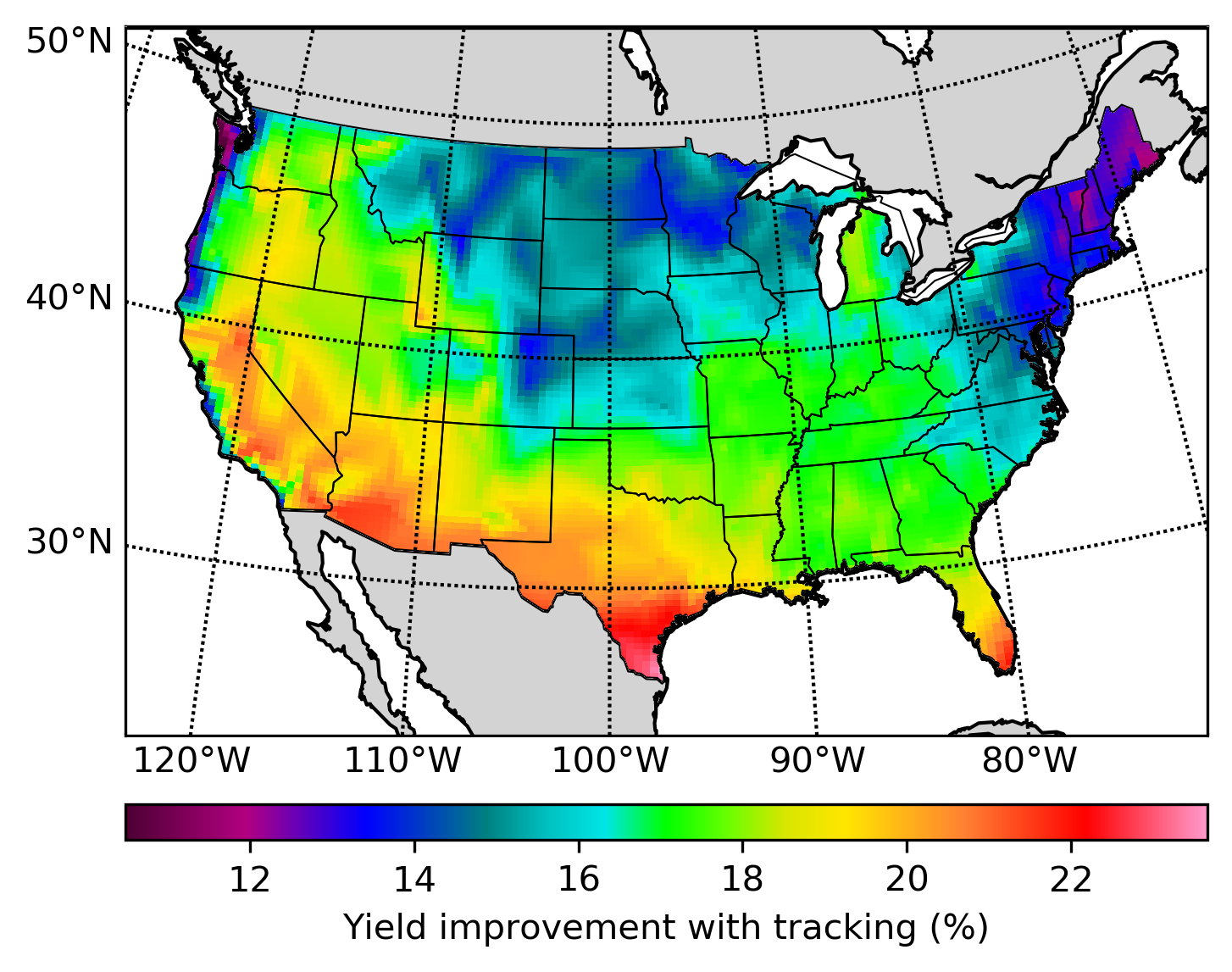}
    \caption{Yearly energy yield improvement factor obtained by mounting silicon PV modules on horizontal single axis trackers rather than at a fixed tilt angle.}
\label{fig:ratio_tracking}
\end{figure}

Among all the utility scale PV systems installed in the US in 2016, 80\% were tracking systems\cite{NREL_2017}. The most common type of PV tracking is currently horizontal single axis (HSAT) tracking. We present in Fig. \ref{fig:ratio_tracking} the ratio of the yearly energy production of silicon based HSAT systems relative to that of fixed tilt systems.

\begin{figure}[h]
\centering
    \includegraphics[width=0.47\textwidth]{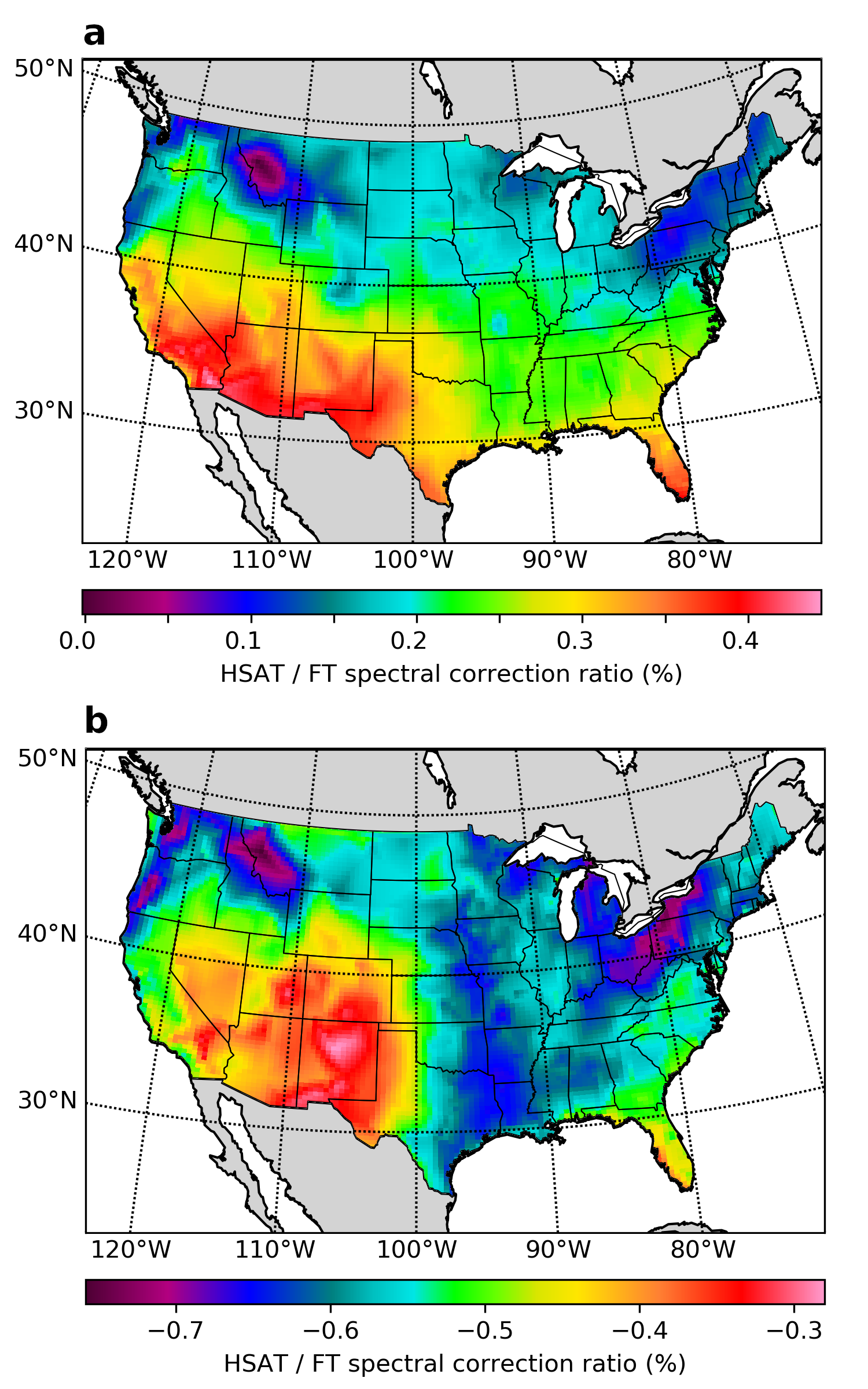}
    \caption{Ratio of spectral correction factors for tracking (HSAT) and fixed tilt (FT) systems. \textbf{a}, Silicon. \textbf{b}, Perovskite.}
\label{fig:tracking2}
\end{figure}

Using tracking to increase the average irradiance on the POA comes at the cost of increased solar cell temperatures. But thermal effects are only slightly detrimental to the efficiency of HSAT systems because their higher yearly energy production (Fig. \ref{fig:ratio_tracking}) is mostly due to a more spread out energy production along the course of each day, and not to significantly higher peak POA irradiances. Furthermore, we find that the spectral correction factors for silicon based tracking systems systems are more favorable that those of fixed tilt systems. The ratio of spectral correction factors for HSAT ($f_{s}^\text{HSAT}$) and fixed tilt systems ($f_{s}^\text{FT}$) is presented as a percentage as $(f_{s}^\text{HSAT}/f_{s}^\text{FT} - 1)$ in Fig. \ref{fig:tracking2}. For silicon systems (Fig. \ref{fig:tracking2}a), tracking is found to be favoured by spectral effects in all of the locations that we have studied, and comparing Fig. \ref{fig:si}a with Fig. \ref{fig:tracking2}a reveals that the spectral correction ratio is most favourable for tracking systems in those areas with the highest yearly energy production. Spectral effects further increase the energy yield advantage of silicon based trackers because trackers collect more sunlight during sunrise and sunset, and during these times the spectrum peaks at lower energies due to the higher air mass. Although this increases losses due to photons with energy lower than the band gap, the effect that prevails is a reduction in carrier thermalization losses, as the band gap of silicon (1.12 eV) is smaller than the optimal band gap for maximum yearly energy production (1.35 eV)\cite{ripalda_solar_2018}. Conversely, if perovskites or other high band gap single junctions are used, spectral effects favor a fixed tilt geometry, as shown in Fig. \ref{fig:tracking2}b. Because the POA irradiance of tracking systems is higher than the POA irradiance of fixed tilt systems, the energy yield is always higher for trackers, but this advantage is reduced in the case of perovskite absorbers due to spectral effects. The capital cost of utility scale PV power plants is 1.03 \$/W for fixed tilt and 1.11 \$/W for tracker systems\cite{NREL_2017}. Because the cost difference is only 7.76\%, spectral effects make the return on investment for silicon based trackers 5.8\% more favorable.

\section*{Multijunctions}
\begin{figure}[ht!]
\centering
    \includegraphics[width=0.47\textwidth]{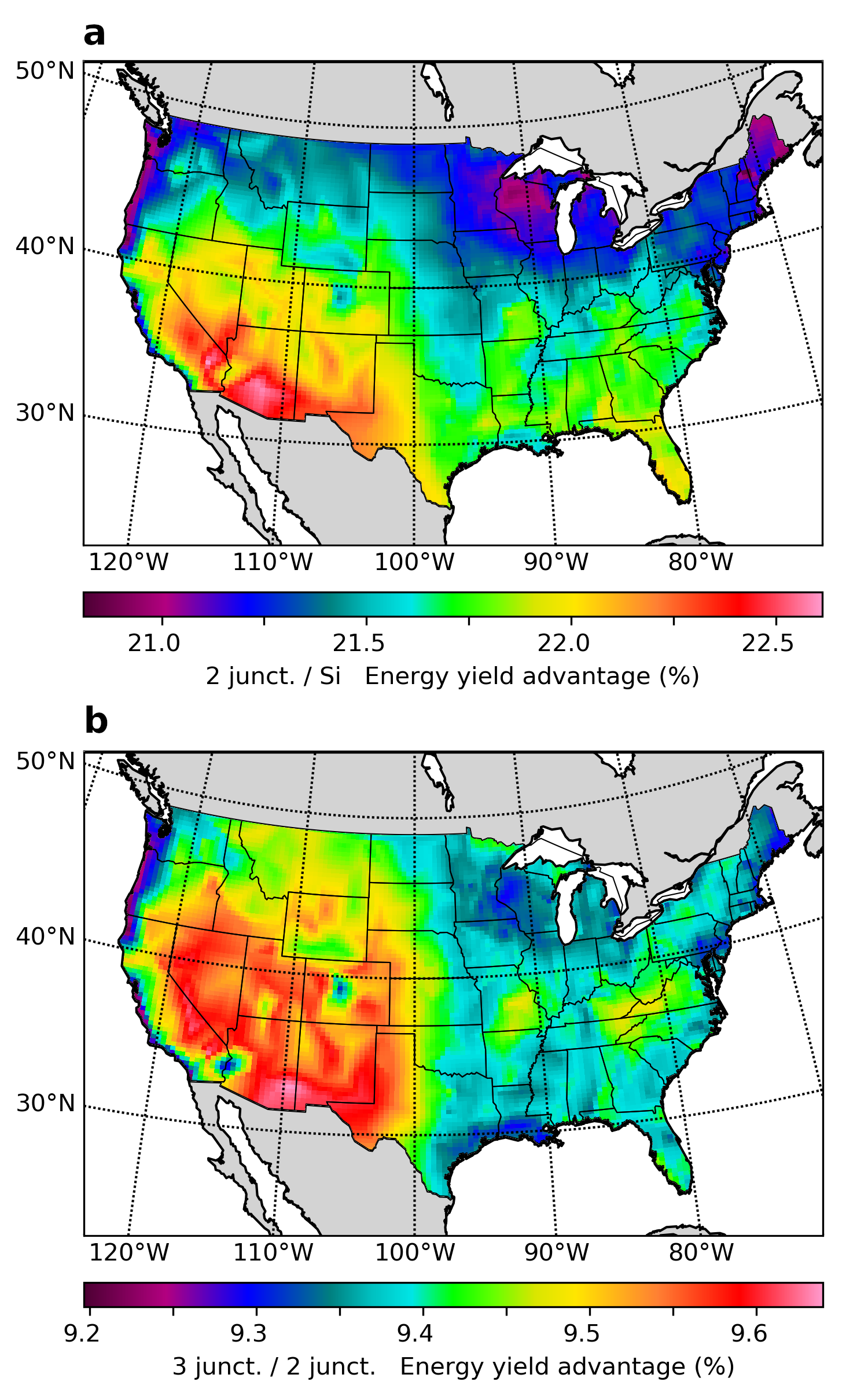}
    \caption{Energy yield ratios as a function of the number of junctions. The POA irradiance is here the same in all cases, so the efficiency ratio is the same as the energy yield ratio. a), Ratio of the energy yield obtained with an optimal double junction to the energy yield of silicon under global irradiance with horizontal single axis tracking. b) Ratio of the energy yield obtained with an optimal triple junction to the energy yield of an optimal double junction under global irradiance with horizontal single axis tracking.}
\label{fig:mj}
\end{figure}

An often raised concern about multijunction technology is its sensitivity to spectral variations. We present in Fig.\ref{fig:mj}a the energy yield ratio of an optimal series connected double junction relative to a silicon single junction under global irradiance with horizontal single axis tracking. In this case the POA irradiance is the same for both systems, and thus the energy yield ratio is the same as the yearly average efficiency ratio. The band gaps of the dual junction here discussed are those found as optimal in our previous work, 1.126 eV and 1.687 eV for the bottom and top junctions respectively\cite{ripalda_solar_2018}.

We find that double junctions are most advantageous in high irradiance locations due to a lower sensitivity to high temperatures. In the south, a 22\% yearly energy yield advantage might provide a market entrance opportunity for dual junction modules, especially in the residential market, where area constraints increase the value of high efficiency systems, and modules only represent about 10\% of the total cost\cite{NREL_2017}, allowing for multijunction module costs three times higher than current module costs. This target might be compatible with a recent cost reduction road map for III-V multijunctions published by NREL\cite{essig_raising_2017}. Alternatively, multijunctions based on perovskites have recently surpassed the efficiency of silicon single junctions\cite{green_solar_2020}, and reported degradation rates are also improving rapidly\cite{hou_efficient_2020, xu_triple-halide_2020}.

The areas most favourable for multijunctions in Fig.\ref{fig:mj}a have a large overlap with the regions most favourable for tracking in Fig. \ref{fig:tracking2}b. This reinforces a synergy between these two technologies given by the fact that the revenue generated by a photovoltaic system results from the product of a number of factors such as solar cell efficiency, inverter efficiency, cell interconnection efficiency, POA irradiance, and transmission of antireflective coatings and encapsulating materials. An increase in any of these factors makes it more profitable to invest in increasing any of the other factors. 

The technical complexity and cost of multijunctions has a super-linear trend with the number of junctions, whereas the efficiency has a sub-linear increase with the number of junctions. So it remains unclear what would be the number of junctions that maximizes the return on investment, partly due to location dependent effects. The relative improvement in the yearly energy production obtained by replacing an optimal double junction with a triple junction is shown in Fig.\ref{fig:mj}b. In both cases the bottom junction is chosen to be silicon, as its band gap is nearly optimal, and it has a high performance/cost ratio. The middle and top junction band gaps for the triple junction are those found as optimal for a silicon based series connected triple junction in our previous work, 1.48, 1.94 eV, respectively\cite{ripalda_solar_2018}. 

As expected, the spectral correction factors for the series connected double and triple junction (not shown) are more adverse than those of the silicon single junction. They follow a geographical pattern closely matching that in Fig.\ref{fig:si}c, suggesting that the spectral sensitivity effects are mostly given by the absorption threshold of the silicon bottom junction. The spectral corrections range from -3.4\% to -1.1\% for the double junction and -4.4\% to -2.1\% for the triple junction. If the photocurrent from the silicon bottom junction is collected separately using a three or four terminal configuration (neglecting cell interconnection losses), the spectral corrections range from -2.2\% to 1.5\% for the double junction and -4.0\% to -0.7\% for the triple junction. So a multi-terminal configuration is most beneficial for the double junction. The energy yield of multi-terminal silicon based tandems has also been recently studied by Schulte-Huxel \textit{et al.}\cite{schulte-huxel_energy_2018}, using technologically relevant but not optimal band gaps (GaAs and GaInP), and Essig \textit{et al.} recently reported record efficiencies with silicon based tandems using these materials \cite{essig_raising_2017}. Using the corresponding band gaps (1.42 eV for GaAs and 1.85 eV for GaInP) with our model, we reproduce the results reported by Schulte-Huxel et al.\cite{schulte-huxel_energy_2018}, obtaining larger gains for the multi-terminal configuration than when using optimal band gaps. Thus the multi-terminal configuration is of most interest when the optimal band gaps for the series connected configuration cannot be used due to technological constraints. This conclusion is also supported by the recent work by Mathews \textit{et al.}\cite{mathews_predicted_2020}.

\begin{figure}[tbh!]
\centering
    \includegraphics[width=0.47\textwidth]{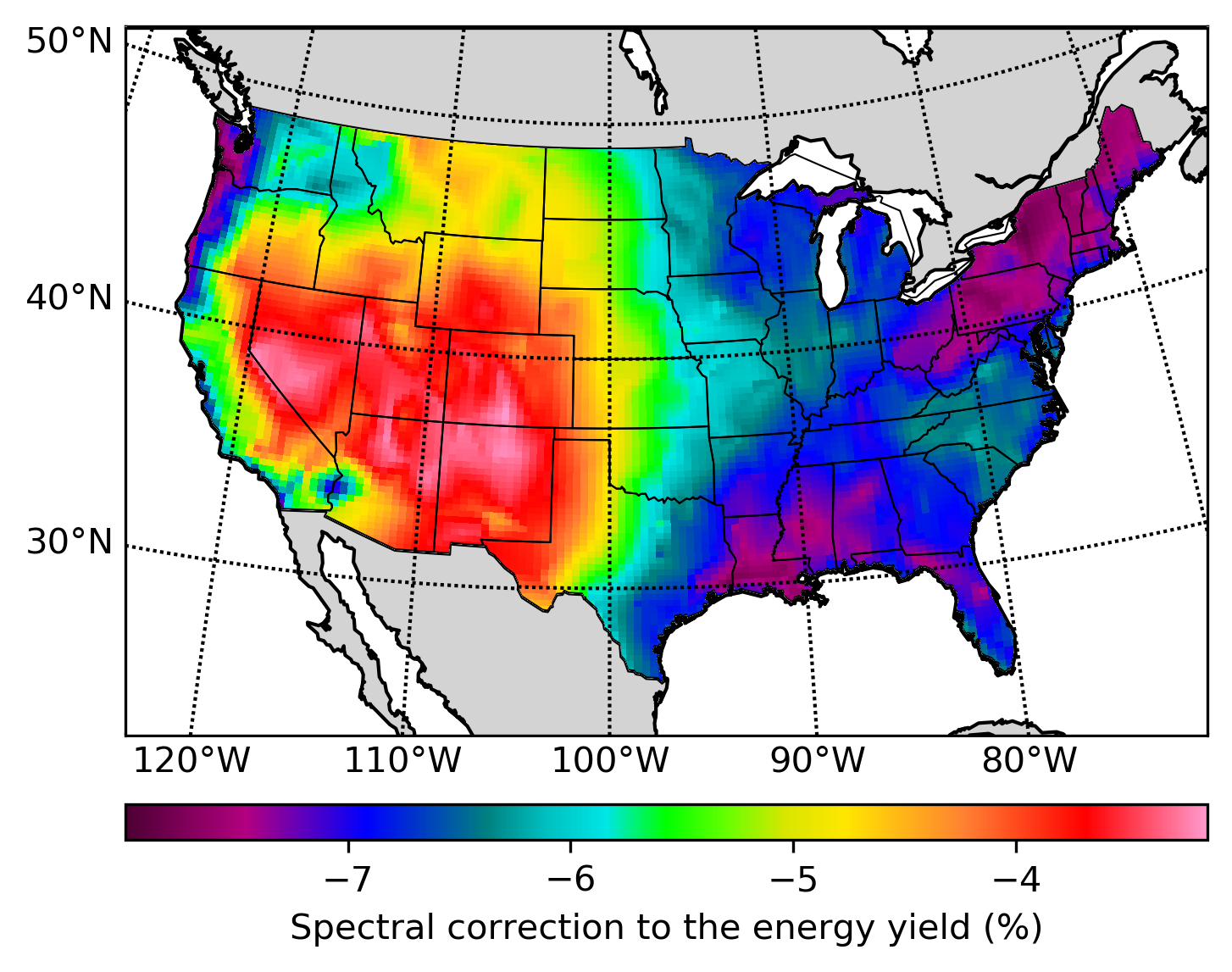}
    \caption{Spectral correction factor for the yearly energy production of a 6 junction solar cell with the band gaps of the current record device under global irradiance\cite{geisz_building_2018}.}
\label{fig:6j}
\end{figure}

As an example of extreme spectral sensitivity, we have considered the case of an optimal series connected 6 junction device under global spectra. The band gaps of the 6 junction architecture here discussed are those of the current record for a solar cell under the global spectrum\cite{geisz_building_2018, green_solar_2020}. The energy yield advantage over silicon single junctions ($E_\text{y}^\text{6j} / E_\text{y}^\text{Si} - 1$) ranges from 50.8\% in the Rocky Mountains to 38.7\% in New England. The spectral correction factor for the 6 junction device is shown in Fig. \ref{fig:6j}.  The geographical pattern is almost the opposite of all the previous cases, with high altitude locations being the least adversely affected by spectral sensitivity effects. This different pattern here suggests that the spectral sensitivity of this device is of a fundamentally different type than in the previous cases. While silicon and silicon based multijunctions have a spectral sensitivity mostly determined by the absorption threshold of silicon, the spectral sensitivity of this 6 junction device is mostly given by the current matching constraint. High altitude reduces losses caused by the atmosphere, reducing spectral variability and current mismatch effects in the 6 junction case.

\section*{Fine tuning for specific locations}
As recently discussed by Parent \textit{et al.}, the energy yield of multijunctions can be increased by optimizing the band gaps using local spectra and meteorological conditions\cite{parent_2019}. We have re-optimized the band gaps of silicon based triple junctions at a few representative locations. In the data presented here (Table \ref{table:fine}) the bottom junction band gap is fixed as we assume a silicon based tandem. If this constrain is relaxed, we find the required bottom junction band gap adjustments to reach a local efficiency maximum are typically small, while top junctions are the ones that require larger adjustments. This is expected because, regardless of geographical location, the efficiency maxima as a function of the bottom junction band gap are mostly given by the atmospheric absorption band thresholds, as discussed by McMahon \textit{et al.}\cite{mcmahon_multijunction_2017}. Higher top junction band gaps are favoured in hot areas as this leads to a reduction of recombination voltage losses, but spectral effects lead to exceptions to this rule, as in the case of the north pacific coast, where high band gaps are favoured due to spectral effects. The obtained efficiency improvement is typically of about 0.5\%.

\renewcommand\tabcolsep{4pt}
\begin{table}[htb]
\sffamily
\small
\caption{\label{table:fine} Fine tuning of band gaps at specific locations for series constrained silicon based triple junctions. The reference efficiency is obtained with middle and top junction band gaps of 1.48 and 1.94 eV, respectively.}
\begin{tabular}{rcccc}
            & \textbf{Mid. gap} & \textbf{Top gap} & \textbf{Ref. Eff.} & \textbf{Eff.} \\
 & \textbf{eV} & \textbf{eV} & \textbf{\%} & \textbf{\%} \\
Leadville, CO  & 1.490           & 1.975        & 34.95          & 35.37     \\
Denver, CO  & 1.494           & 1.978        & 34.81          & 35.23     \\
Mojave, CA  & 1.499           & 1.981        & 34.70          & 35.26     \\
Tucson, AZ  & 1.500           & 1.986        & 34.33          & 34.95     \\
Astoria, OR & 1.504           & 1.987        & 35.40          & 36.16
\end{tabular}
\end{table}

\section*{Uncertainty vs. number of spectra}
Our results suggest that forecasting the yearly energy production of PV systems requires location specific solar spectra. Yearly spectral sets with thousands of spectra per year and location are available from the National Solar Radiation Database (NSRDB)\cite{NSRDB_2018, xie_fast_2018, xie_fast_2019}. The dataset used in this work comprises 16 million spectra, each with 2002 wavelengths and associated meteorological data. The number of required spectra can be reduced using statistical techniques such as binning\cite{garcia_spectral_2018}, and machine learning clustering\cite{ripalda_solar_2018}. Here we cluster the spectra according not only to their spectral content as in our previous work\cite{ripalda_solar_2018}, but also according to other correlated meteorological data such as wind speed and ambient temperature, as these also have an effect on PV efficiency. 

In the previous sections we have used 18 clustered spectra per location, as we have previously determined that this leads to an uncertainty in the results typically smaller than 0.2\% while still reducing the computational cost by several orders of magnitude\cite{ripalda_solar_2018}. 

\begin{figure}[tbh!]
\centering
    \includegraphics[width=0.47\textwidth]{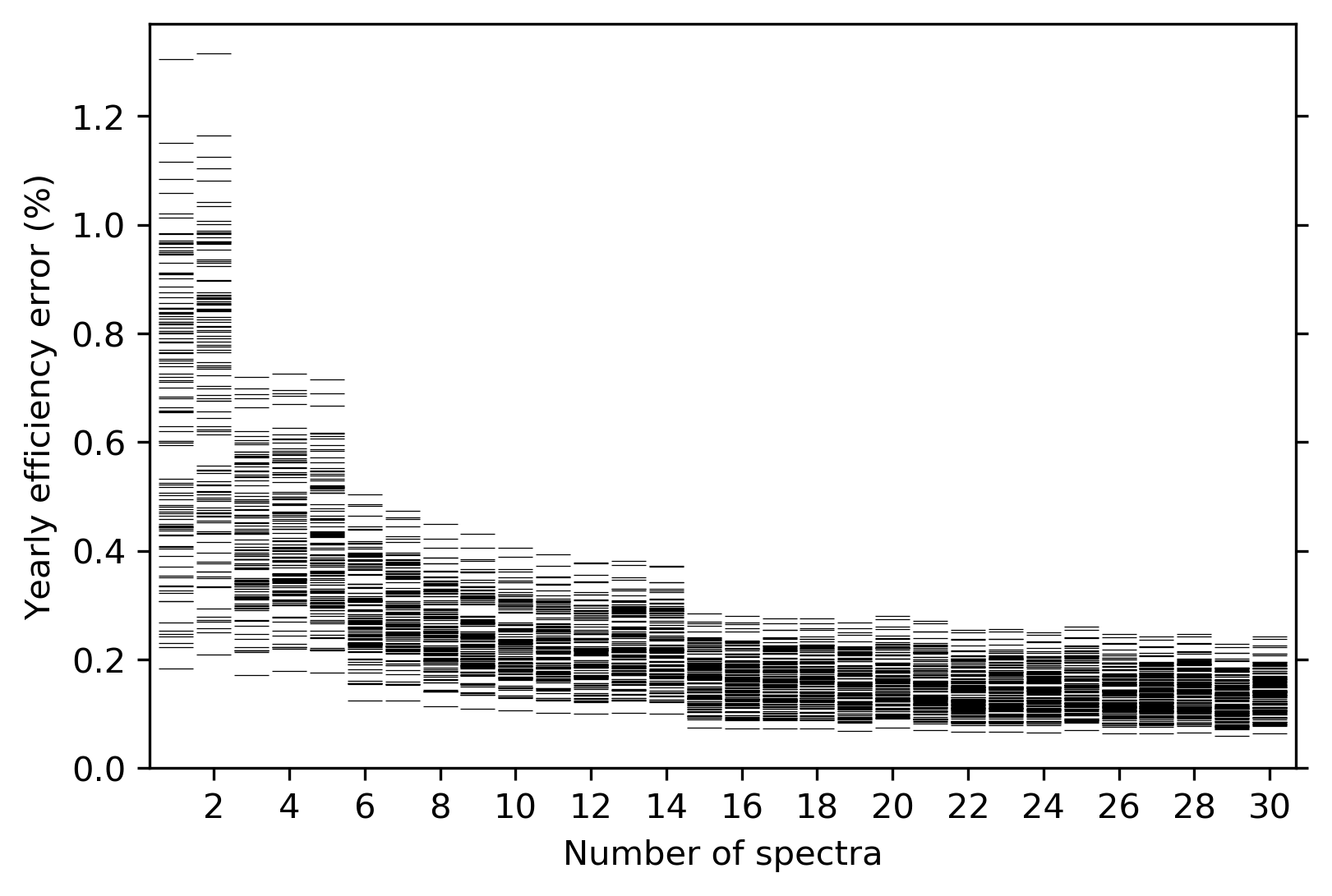}
    \caption{Convergence of the yearly averaged efficiency as a function of the number of clustered spectra. This example corresponds to a set of 140 triple junctions with random but nearly optimal band gaps (efficiency within 2\% of the maximum) on a horizontal single axis tracker at Elko, Nevada.}
\label{fig:convergence}
\end{figure}

In this section we study how the quality of the obtained results improves as the number of spectra is increased, using as a reference the results obtained with the whole data-set. In Fig. \ref{fig:convergence} we show the efficiency error statistics as a function of the number of spectra for triple junction modules on horizontal single axis trackers. As shown in Fig. \ref{fig:convergence}, there is little benefit obtained by increasing the number of proxy spectra beyond 20, and the uncertainty in energy yield forecasts is likely to be dominated by other factors such as the uncertainties on the spectrally integrated irradiance, module degradation, soiling rates, and other loss mechanisms at the module and system level that are out of the scope of this work. 

\section*{Discussion}
The contiguous United States spans a wide range of latitudes and atmospheric conditions. As a consequence, the yearly averaged photovoltaic efficiency of silicon modules can vary with location by up to 1.4\% absolute efficiency. Spectral sensitivity effects account for about half of the geographical variability of efficiencies, with spectral correction factors ranging from -2.0\% to 1.1\% in terms of the energy yield, or -0.5\% to 0.3\% in terms of absolute efficiency. We find that thermal effects predominate over spectral sensitivity, and slightly reduce the economic advantage of high irradiance locations.  While the former are mostly determined by latitude and irradiance, the latter are mostly determined by topographical altitude and atmospheric phenomena. Spectral sensitivity effects are found to favour silicon horizontal single axis trackers over fixed tilt systems by up to 0.45\% of the energy yield in high irradiance locations at lower latitudes, but a fixed tilt geometry is favored instead for perovskites and other high band gap absorbers such as CdTe. The energy yield loss caused by spectral sensitivity in multijunctions is found to be roughly proportional to the number of junctions. If the optimal band gaps are used, this loss is not significantly mitigated by using a multi-terminal configuration for current extraction. Nevertheless, because the choice of band gaps is often constrained, multi-terminal configurations greatly enhance the flexibility of multijunction design. The data here presented clearly shows that location specific solar spectra are required for accurate predictions of the energy yield, but rather than sets of thousands of spectra for each location covering a whole year, it suffices to use a few characteristic spectra for each location. A possible future line of research, inspired by the recent work by Warmann and Atwater\cite{warmann_predicting_2019}, is to identify a set of geographical, atmospheric and meteorological parameters that correctly predicts the yearly energy yield in all locations within an acceptable uncertainty margin. But the most pressing need at the moment seems to be to decrease the uncertainty margin in the spectral irradiance data used as input in models such as the one here discussed. Such advances should lead to significantly reduced uncertainty in PV production forecasts, and consequently lower risk and financial cost for PV projects.

\section*{Methods}
\small
\textbf{Solar cell model and proxy spectra.} We have used a modified detailed balance method with a set of proxy spectra obtained from clustering of yearly spectral sets as described in Ref. \cite{ripalda_solar_2018}. Building on state of the art understanding of fundamental losses\cite{hirst_fundamental_2011}, our model is based on previous work by Geisz \textit{et al.} and García \textit{et al.}. \cite{geisz_generalized_2015, garcia_spectral_2018}. All the parameters in the model are the same as in our previous work with the exception of the external radiative efficiency for CdTe, where we optimistically assumed 0.01\% rather than the default value of 1\%. A parameter sensitivity analysis including all relevant parameters in our model can be found in Ref. \cite{chicago_2019}.  The global spectral irradiance data was downloaded from the National Solar Radiation Database (NSRDB) as derived from the FARMS-NIT model valid for all sky conditions, including the effects of clouds\cite{xie_fast_2018, xie_fast_2019}. The used spectra and meteorological data are historical data from 2017. The spectral irradiance is modified by an angle of incidence modifier accounting for increased reflectance at grazing angles. We have used the physically based angle of incidence modifier implemented in the PVLIB python open source library \cite{f_holmgren_pvlib_2018, de_soto_improvement_2006}.

The initial data set includes 16 million spectra (8760 spectra for 913 locations and 2 different collection geometries). These are reduced to 18 characteristic spectra per location used as a proxy for the whole yearly data-set. The proxy spectra are obtained by machine learning clustering of the initial data set\cite{ripalda_solar_2018}. The band gaps were optimized for maximum yearly energy yield as described in Ref. \cite{ripalda_solar_2018}. 

\textbf{Temperature model.} We obtain solar cell temperatures from the ambient temperature, the irradiance, and the wind speed using the PVLIB python open source library implementation of the empirically fitted Sandia photovoltaic array performance model using the default parameters for a polymer-back open rack array \cite{f_holmgren_pvlib_2018}. To include the effect of efficiency on solar cell temperatures, the irradiance values used as input in the Sandia temperature model are modified by a thermal correction factor that is a function of the efficiency, as the efficiency of the modules used to fit the Sandia temperature model was rather low ($\eta_0$ = 16.5\%), and the power carried away as electricity does not contribute to raise the solar cell temperature. This factor is given by $\left ( 1 - R - \eta \right )/\left ( 1 - R - \eta_0 \right )$, where $\eta$ is the efficiency, and $R$ is the reflectivity. The efficiency values used to calculate the thermal correction factor are those in Table 1 of Ref. \cite{ripalda_solar_2018} to avoid the need for self consistent iteration, as the temperature would depend on the efficiency and vice versa. The value assumed for $R$ is arbitrarily chosen to be 0.05, but the sensitivity of the results to both $R$ or $\eta$ is small (doubling the value assumed for $R$ or increasing the efficiency by 5\% reduces the resulting temperature by 0.5$^\circ$C and this increases the efficiency of a silicon single junction by 0.03\%).

\textbf{Caveats and shortcomings.} Although out of the scope of the present work, in practice there are other important effects on the energy production and return on investment of PV installations, such as the higher degradation rate with increased module temperatures \cite{ascencio-vasquez_global_2019, vazquez_reliability_2017}, and the geographical and seasonal variations of soiling rates\cite{garcia_soiling_2011, sarver_comprehensive_2013, ilse_techno-economic_2019}, as well as other loss mechanisms at the module and system level. When PV module temperatures are above a certain threshold, it becomes economically advantageous to turn trackers away from the sun, reducing the risk of damage but also the energy production. Similarly, the frequency of cleaning is also determined by a balance between the energy yield and operating costs. Both effects are specially adverse in arid regions with high temperatures and infrequent rain. We reduce the reported specific energy yields by 3\% due to shadow losses, 2\% due to inverter clipping and inverter efficiency losses, 2\% due to system degradation and failures, 2\% due to soiling, and 1\% due to other effects such as DC and AC wire losses and mismatch losses, among others. We do not attempt to do a location dependent bottom-up model of system level losses, as these effects have little correlation with the spectral effects that are the main subject of this work.

Ground based observations of the solar spectra are available from a limited number of locations, covering a limited time span. Therefore a consistent large data-set covering large regions and time-spans can only be obtained from modelled data based on satellite images, and later validated by comparison with spectra measured from the ground\cite{NREL_2015}. The NREL-NSRDB spectra used in this work have been derived from the FARMS-NIT model and validated in comparison with measured spectra from the NREL Solar Radiation Research Laboratory in Ref. \cite{NREL_Xie_2019}. These ground based measurements unavoidably have their own uncertainties and biases, and thus complete agreement with physics based models cannot be expected. The FARMS-NIT data in comparison with surface based observations have shown a percentage error in the 1.86 \% to 2.28\% range, while the previous model from NREL (TMYSPEC) had percentage errors ranging from -3.47\% to -16.27\% \cite{NREL_Xie_2019}. Although this represents a large improvement over the previous state of the art, there is still a clear need to reduce these uncertainties to improve the bankability of PV systems. The potential economic return of such advances clearly justifies the required research effort. In the mean time, the qualitative trends here revealed illustrate phenomena that need to be accounted for in order to improve photovoltaic energy production forecasts.

The spectral irradiance data for horizontal single axis tracker systems provided by the NSRDB is obtained in the limit of low ground cover ratios without back-tracking\cite{NREL_Xie_2019}. Systems with high ground cover ratios can be expected to be less sensitive to the changes in the spectra during the early morning and late afternoon.

\section*{Data and code availability}
The data and code required to reproduce the results here presented are available as open source at \url{https://github.com/Ripalda/Tandems}

\section*{Acknowledgments}
We gratefully acknowledge the scientific and technical input from Dr. J. Buencuerpo. This work would have not been possible without the data and computing resources made publicly available by NREL as part of the NSRDB. Funding was provided by the Spanish Government and the European Union through MCIU-AEI-FEDER-UE (ENE2017‐91092‐EXP, RTI2018-096937-B-C22, RYC-2017-21995) and Comunidad de Madrid (P2018/EMT-4308). D. Chemisana thanks "Institució Catalana de Recerca i Estudis Avançats (ICREA)" for the ICREA Acadèmia award. I.G. is funded by Ministerio de Economía y Competitividad through the Ramón y Cajal program (RYC-2014-15621).

\section*{Author contributions}
J.M.R. wrote the initial version of the manuscript and python code. D.C. designed an initial version of this study and provided much of the motivation for this work. J.M.L. contributed to the methodology and reviewed the python code. I.G. contributed to the methodology (proxy spectra, physics of multijunction devices). All authors jointly discussed the results and contributed to the manuscript.

\section*{Competing interests}
The authors declare no competing interests.







%

\small
\bibliography{tand}

\begin{thebibliography}{10}
\expandafter\ifx\csname url\endcsname\relax
  \def\url#1{\texttt{#1}}\fi
\expandafter\ifx\csname urlprefix\endcsname\relax\def\urlprefix{URL }\fi
\providecommand{\bibinfo}[2]{#2}
\providecommand{\eprint}[2][]{\url{#2}}

\bibitem{science_terawatt_2019}
\bibinfo{author}{Haegel, N.~M.} \emph{et~al.}
\newblock \bibinfo{title}{Terawatt-scale photovoltaics: {Transform} global
  energy}.
\newblock \emph{\bibinfo{journal}{Science}} \textbf{\bibinfo{volume}{364}},
  \bibinfo{pages}{836--838} (\bibinfo{year}{2019}).
\newblock
  \urlprefix\url{http://www.sciencemag.org/lookup/doi/10.1126/science.aaw1845}.

\bibitem{kurtz_difference_1991}
\bibinfo{author}{Kurtz, S.~R.}, \bibinfo{author}{Olson, J.} \&
  \bibinfo{author}{Faine, P.}
\newblock \bibinfo{title}{The difference between standard and average
  efficiencies of multijunction compared with single-junction concentrator
  cells}.
\newblock \emph{\bibinfo{journal}{Solar Cells}} \textbf{\bibinfo{volume}{30}},
  \bibinfo{pages}{501--513} (\bibinfo{year}{1991}).
\newblock
  \urlprefix\url{https://linkinghub.elsevier.com/retrieve/pii/037967879190081Y}.

\bibitem{faine_influence_1991}
\bibinfo{author}{Faine, P.}, \bibinfo{author}{Kurtz, S.~R.},
  \bibinfo{author}{Riordan, C.} \& \bibinfo{author}{Olson, J.}
\newblock \bibinfo{title}{The influence of spectral solar irradiance variations
  on the performance of selected single-junction and multijunction solar
  cells}.
\newblock \emph{\bibinfo{journal}{Solar Cells}} \textbf{\bibinfo{volume}{31}},
  \bibinfo{pages}{259--278} (\bibinfo{year}{1991}).
\newblock
  \urlprefix\url{https://linkinghub.elsevier.com/retrieve/pii/037967879190027M}.

\bibitem{kurtz_projected_1997}
\bibinfo{author}{Kurtz, S.}, \bibinfo{author}{Myers, D.} \&
  \bibinfo{author}{Olson, J.}
\newblock \bibinfo{title}{Projected performance of three- and four-junction
  devices using {GaAs} and {GaInP}}.
\newblock In \emph{\bibinfo{booktitle}{PVSC 26 Conference Record}},
  \bibinfo{pages}{875--878} (\bibinfo{publisher}{IEEE},
  \bibinfo{address}{Anaheim, CA, USA}, \bibinfo{year}{1997}).
\newblock \urlprefix\url{http://ieeexplore.ieee.org/document/654226/}.

\bibitem{chan_impact_2014}
\bibinfo{author}{Chan, N. L.~A.}, \bibinfo{author}{Brindley, H.~E.} \&
  \bibinfo{author}{Ekins-Daukes, N.~J.}
\newblock \bibinfo{title}{Impact of individual atmospheric parameters on {CPV}
  system power, energy yield and cost of energy: {Impact} of individual
  atmospheric parameters on {CPV} systems}.
\newblock \emph{\bibinfo{journal}{Progress in Photovoltaics: Research and
  Applications}} \textbf{\bibinfo{volume}{22}}, \bibinfo{pages}{1080--1095}
  (\bibinfo{year}{2014}).
\newblock \urlprefix\url{http://doi.wiley.com/10.1002/pip.2376}.

\bibitem{ekins-daukes_brighten_2019}
\bibinfo{author}{Ekins-Daukes, N.} \& \bibinfo{author}{Kay, M.}
\newblock \bibinfo{title}{Brighten the dark skies}.
\newblock \emph{\bibinfo{journal}{Nature Energy}} \textbf{\bibinfo{volume}{4}},
  \bibinfo{pages}{633--634} (\bibinfo{year}{2019}).
\newblock \urlprefix\url{http://www.nature.com/articles/s41560-019-0440-0}.

\bibitem{sweerts_estimation_2019}
\bibinfo{author}{Sweerts, B.} \emph{et~al.}
\newblock \bibinfo{title}{Estimation of losses in solar energy production from
  air pollution in {China} since 1960 using surface radiation data}.
\newblock \emph{\bibinfo{journal}{Nature Energy}} \textbf{\bibinfo{volume}{4}},
  \bibinfo{pages}{657--663} (\bibinfo{year}{2019}).
\newblock \urlprefix\url{http://www.nature.com/articles/s41560-019-0412-4}.

\bibitem{xie_fast_2018}
\bibinfo{author}{Xie, Y.} \& \bibinfo{author}{Sengupta, M.}
\newblock \bibinfo{title}{A fast all-sky radiation model for solar applications
  with narrowband irradiances on tilted surfaces ({FARMS}-{NIT}): Part i. the
  clear-sky model}.
\newblock \emph{\bibinfo{journal}{Solar Energy}}
  \textbf{\bibinfo{volume}{174}}, \bibinfo{pages}{691--702}
  (\bibinfo{year}{2018}).
\newblock
  \urlprefix\url{https://linkinghub.elsevier.com/retrieve/pii/S0038092X18309502}.

\bibitem{xie_fast_2019}
\bibinfo{author}{Xie, Y.}, \bibinfo{author}{Sengupta, M.} \&
  \bibinfo{author}{Wang, C.}
\newblock \bibinfo{title}{A fast all-sky radiation model for solar applications
  with narrowband irradiances on tilted surfaces ({FARMS}-{NIT}): Part {II}.
  the cloudy-sky model}.
\newblock \emph{\bibinfo{journal}{Solar Energy}}
  \textbf{\bibinfo{volume}{188}}, \bibinfo{pages}{799--812}
  (\bibinfo{year}{2019}).
\newblock
  \urlprefix\url{https://linkinghub.elsevier.com/retrieve/pii/S0038092X19306334}.

\bibitem{NSRDB_2018}
\bibinfo{author}{Sengupta, M.} \emph{et~al.}
\newblock \bibinfo{title}{The {National} {Solar} {Radiation} {Data} {Base}
  ({NSRDB})}.
\newblock \emph{\bibinfo{journal}{Renewable and Sustainable Energy Reviews}}
  \textbf{\bibinfo{volume}{89}}, \bibinfo{pages}{51--60}
  (\bibinfo{year}{2018}).

\bibitem{ripalda_solar_2018}
\bibinfo{author}{Ripalda, J.~M.}, \bibinfo{author}{Buencuerpo, J.} \&
  \bibinfo{author}{García, I.}
\newblock \bibinfo{title}{Solar cell designs by maximizing energy production
  based on machine learning clustering of spectral variations}.
\newblock \emph{\bibinfo{journal}{Nature Communications}}
  \textbf{\bibinfo{volume}{9}}, \bibinfo{pages}{5126} (\bibinfo{year}{2018}).
\newblock \urlprefix\url{http://www.nature.com/articles/s41467-018-07431-3}.

\bibitem{vossier_is_2017}
\bibinfo{author}{Vossier, A.}, \bibinfo{author}{Riverola, A.},
  \bibinfo{author}{Chemisana, D.}, \bibinfo{author}{Dollet, A.} \&
  \bibinfo{author}{Gueymard, C.~A.}
\newblock \bibinfo{title}{Is conversion efficiency still relevant to qualify
  advanced multi-junction solar cells?: {Is} efficiency relevant with advanced
  {MJ} cells?}
\newblock \emph{\bibinfo{journal}{Progress in Photovoltaics: Research and
  Applications}} \textbf{\bibinfo{volume}{25}}, \bibinfo{pages}{242--254}
  (\bibinfo{year}{2017}).

\bibitem{garcia_spectral_2018}
\bibinfo{author}{Garcia, I.} \emph{et~al.}
\newblock \bibinfo{title}{Spectral binning for energy production calculations
  and multijunction solar cell design}.
\newblock \emph{\bibinfo{journal}{Progress in Photovoltaics: Research and
  Applications}} \textbf{\bibinfo{volume}{26}}, \bibinfo{pages}{48--54}
  (\bibinfo{year}{2018}).

\bibitem{dirnberger}
\bibinfo{author}{Dirnberger, D.}, \bibinfo{author}{Blackburn, G.},
  \bibinfo{author}{Müller, B.} \& \bibinfo{author}{Reise, C.}
\newblock \bibinfo{title}{On the impact of solar spectral irradiance on the
  yield of different pv technologies}.
\newblock \emph{\bibinfo{journal}{Solar Energy Materials and Solar Cells}}
  \textbf{\bibinfo{volume}{132}}, \bibinfo{pages}{431 -- 442}
  (\bibinfo{year}{2015}).
\newblock
  \urlprefix\url{http://www.sciencedirect.com/science/article/pii/S0927024814005169}.

\bibitem{edu}
\bibinfo{author}{Fernández, E.~F.}, \bibinfo{author}{Almonacid, F.},
  \bibinfo{author}{Ruiz-Arias, J.} \& \bibinfo{author}{Soria-Moya, A.}
\newblock \bibinfo{title}{Analysis of the spectral variations on the
  performance of high concentrator photovoltaic modules operating under
  different real climate conditions}.
\newblock \emph{\bibinfo{journal}{Solar Energy Materials and Solar Cells}}
  \textbf{\bibinfo{volume}{127}}, \bibinfo{pages}{179 -- 187}
  (\bibinfo{year}{2014}).
\newblock
  \urlprefix\url{http://www.sciencedirect.com/science/article/pii/S0927024814002347}.

\bibitem{pvmaps}
\bibinfo{author}{Huld, T.}
\newblock \bibinfo{title}{Pvmaps: Software tools and data for the estimation of
  solar radiation and photovoltaic module performance over large geographical
  areas}.
\newblock \emph{\bibinfo{journal}{Solar Energy}}
  \textbf{\bibinfo{volume}{142}}, \bibinfo{pages}{171 -- 181}
  (\bibinfo{year}{2017}).
\newblock
  \urlprefix\url{http://www.sciencedirect.com/science/article/pii/S0038092X16306089}.

\bibitem{kinsey}
\bibinfo{author}{{Kinsey}, G.~S.}
\newblock \bibinfo{title}{Spectrum sensitivity, energy yield, and revenue
  prediction of pv modules}.
\newblock \emph{\bibinfo{journal}{IEEE Journal of Photovoltaics}}
  \textbf{\bibinfo{volume}{5}}, \bibinfo{pages}{258--262}
  (\bibinfo{year}{2015}).

\bibitem{huld_estimating_2015}
\bibinfo{author}{Huld, T.} \& \bibinfo{author}{Gracia-Amillo, A.}
\newblock \bibinfo{title}{Estimating {PV} {Module} {Performance} over {Large}
  {Geographical} {Regions}: {The} {Role} of {Irradiance}, {Air} {Temperature},
  {Wind} {Speed} and {Solar} {Spectrum}}.
\newblock \emph{\bibinfo{journal}{Energies}} \textbf{\bibinfo{volume}{8}},
  \bibinfo{pages}{5159--5181} (\bibinfo{year}{2015}).
\newblock \urlprefix\url{http://www.mdpi.com/1996-1073/8/6/5159}.

\bibitem{lindsay_errors_2020}
\bibinfo{author}{Lindsay, N.}, \bibinfo{author}{Libois, Q.},
  \bibinfo{author}{Badosa, J.}, \bibinfo{author}{Migan-Dubois, A.} \&
  \bibinfo{author}{Bourdin, V.}
\newblock \bibinfo{title}{Errors in {PV} power modelling due to the lack of
  spectral and angular details of solar irradiance inputs}.
\newblock \emph{\bibinfo{journal}{Solar Energy}}
  \textbf{\bibinfo{volume}{197}}, \bibinfo{pages}{266--278}
  (\bibinfo{year}{2020}).
\newblock
  \urlprefix\url{https://linkinghub.elsevier.com/retrieve/pii/S0038092X19312563}.

\bibitem{peters_global_2018}
\bibinfo{author}{Peters, I.~M.}, \bibinfo{author}{Liu, H.},
  \bibinfo{author}{Reindl, T.} \& \bibinfo{author}{Buonassisi, T.}
\newblock \bibinfo{title}{Global {Prediction} of {Photovoltaic} {Field}
  {Performance} {Differences} {Using} {Open}-{Source} {Satellite} {Data}}.
\newblock \emph{\bibinfo{journal}{Joule}} \textbf{\bibinfo{volume}{2}},
  \bibinfo{pages}{307--322} (\bibinfo{year}{2018}).
\newblock
  \urlprefix\url{https://linkinghub.elsevier.com/retrieve/pii/S2542435117301836}.

\bibitem{warmann_predicting_2019}
\bibinfo{author}{Warmann, E.~C.} \& \bibinfo{author}{Atwater, H.~A.}
\newblock \bibinfo{title}{Predicting {Geographic} {Energy} {Production} for
  {Tandem} {PV} {Designs} {Using} a {Compact} {Set} of {Spectra} {Correlated}
  by {Irradiance}}.
\newblock \emph{\bibinfo{journal}{IEEE Journal of Photovoltaics}}
  \textbf{\bibinfo{volume}{9}}, \bibinfo{pages}{1596--1601}
  (\bibinfo{year}{2019}).
\newblock \urlprefix\url{https://ieeexplore.ieee.org/document/8827709/}.

\bibitem{sandia_2004}
\bibinfo{author}{King, D.}, \bibinfo{author}{Boyson, W.} \&
  \bibinfo{author}{Kratochvill, J.}
\newblock \bibinfo{title}{Photovoltaic array performance model}.
\newblock \bibinfo{type}{Tech. Rep.}, \bibinfo{institution}{Sandia Report
  SAND2004-3535} (\bibinfo{year}{2004}).

\bibitem{dupre_physics_2015}
\bibinfo{author}{Dupré, O.}, \bibinfo{author}{Vaillon, R.} \&
  \bibinfo{author}{Green, M.}
\newblock \bibinfo{title}{Physics of the temperature coefficients of solar
  cells}.
\newblock \emph{\bibinfo{journal}{Solar Energy Materials and Solar Cells}}
  \textbf{\bibinfo{volume}{140}}, \bibinfo{pages}{92--100}
  (\bibinfo{year}{2015}).
\newblock
  \urlprefix\url{https://linkinghub.elsevier.com/retrieve/pii/S0927024815001403}.

\bibitem{green_thermal_2017}
\bibinfo{author}{Dupré, O.}, \bibinfo{author}{Vaillon, R.} \&
  \bibinfo{author}{Green, M.~A.}
\newblock \emph{\bibinfo{title}{Thermal Behavior of Photovoltaic Devices}}
  (\bibinfo{publisher}{Springer}, \bibinfo{year}{2017}).

\bibitem{jacobson_world_2018}
\bibinfo{author}{Jacobson, M.~Z.} \& \bibinfo{author}{Jadhav, V.}
\newblock \bibinfo{title}{World estimates of {PV} optimal tilt angles and
  ratios of sunlight incident upon tilted and tracked {PV} panels relative to
  horizontal panels}.
\newblock \emph{\bibinfo{journal}{Solar Energy}}
  \textbf{\bibinfo{volume}{169}}, \bibinfo{pages}{55--66}
  (\bibinfo{year}{2018}).
\newblock
  \urlprefix\url{https://linkinghub.elsevier.com/retrieve/pii/S0038092X1830375X}.

\bibitem{jung_efficient_2019}
\bibinfo{author}{Jung, E.~H.} \emph{et~al.}
\newblock \bibinfo{title}{Efficient, stable and scalable perovskite solar cells
  using poly(3-hexylthiophene)}.
\newblock \emph{\bibinfo{journal}{Nature}} \textbf{\bibinfo{volume}{567}},
  \bibinfo{pages}{511--515} (\bibinfo{year}{2019}).
\newblock \urlprefix\url{http://www.nature.com/articles/s41586-019-1036-3}.

\bibitem{tress_performance_2019}
\bibinfo{author}{Tress, W.} \emph{et~al.}
\newblock \bibinfo{title}{Performance of perovskite solar cells under simulated
  temperature-illumination real-world operating conditions}.
\newblock \emph{\bibinfo{journal}{Nature Energy}} \textbf{\bibinfo{volume}{4}},
  \bibinfo{pages}{568--574} (\bibinfo{year}{2019}).
\newblock \urlprefix\url{http://www.nature.com/articles/s41560-019-0400-8}.

\bibitem{NREL_2017}
\bibinfo{author}{Fu, R.}, \bibinfo{author}{Feldman, D.},
  \bibinfo{author}{Margolis, R.}, \bibinfo{author}{Woodhouse, M.} \&
  \bibinfo{author}{Ardani, K.}
\newblock \bibinfo{title}{{U.S.} solar photovoltaic system cost benchmark: Q1
  2017}.
\newblock \bibinfo{type}{Tech. Rep.} \bibinfo{number}{NREL/TP-6A20-68925},
  \bibinfo{institution}{NREL} (\bibinfo{year}{2017}).

\bibitem{essig_raising_2017}
\bibinfo{author}{Essig, S.} \emph{et~al.}
\newblock \bibinfo{title}{Raising the one-sun conversion efficiency of
  {III}–{V}/{Si} solar cells to 32.8\% for two junctions and 35.9\% for three
  junctions}.
\newblock \emph{\bibinfo{journal}{Nature Energy}} \textbf{\bibinfo{volume}{2}},
  \bibinfo{pages}{17144} (\bibinfo{year}{2017}).

\bibitem{green_solar_2020}
\bibinfo{author}{Green, M.~A.} \emph{et~al.}
\newblock \bibinfo{title}{Solar cell efficiency tables (version 55)}.
\newblock \emph{\bibinfo{journal}{Progress in Photovoltaics: Research and
  Applications}} \textbf{\bibinfo{volume}{28}}, \bibinfo{pages}{3--15}
  (\bibinfo{year}{2020}).
\newblock
  \urlprefix\url{https://onlinelibrary.wiley.com/doi/abs/10.1002/pip.3228}.

\bibitem{hou_efficient_2020}
\bibinfo{author}{Hou, Y.} \emph{et~al.}
\newblock \bibinfo{title}{Efficient tandem solar cells with solution-processed
  perovskite on textured crystalline silicon}.
\newblock \emph{\bibinfo{journal}{Science}} \textbf{\bibinfo{volume}{367}},
  \bibinfo{pages}{1135--1140} (\bibinfo{year}{2020}).
\newblock
  \urlprefix\url{https://www.sciencemag.org/lookup/doi/10.1126/science.aaz3691}.

\bibitem{xu_triple-halide_2020}
\bibinfo{author}{Xu, J.} \emph{et~al.}
\newblock \bibinfo{title}{Triple-halide wide–band gap perovskites with
  suppressed phase segregation for efficient tandems}.
\newblock \emph{\bibinfo{journal}{Science}} \textbf{\bibinfo{volume}{367}},
  \bibinfo{pages}{1097--1104} (\bibinfo{year}{2020}).
\newblock
  \urlprefix\url{https://www.sciencemag.org/lookup/doi/10.1126/science.aaz5074}.

\bibitem{schulte-huxel_energy_2018}
\bibinfo{author}{Schulte-Huxel, H.}, \bibinfo{author}{Silverman, T.~J.},
  \bibinfo{author}{Deceglie, M.~G.}, \bibinfo{author}{Friedman, D.~J.} \&
  \bibinfo{author}{Tamboli, A.~C.}
\newblock \bibinfo{title}{Energy yield analysis of multiterminal si-based
  tandem solar cells}.
\newblock \emph{\bibinfo{journal}{{IEEE} Journal of Photovoltaics}}
  \textbf{\bibinfo{volume}{8}}, \bibinfo{pages}{1376--1383}
  (\bibinfo{year}{2018}).
\newblock \urlprefix\url{https://ieeexplore.ieee.org/document/8405753/}.

\bibitem{mathews_predicted_2020}
\bibinfo{author}{Mathews, I.}, \bibinfo{author}{Lei, S.} \&
  \bibinfo{author}{Frizzell, R.}
\newblock \bibinfo{title}{Predicted annual energy yield of {III}-v/c-si tandem
  solar cells: modelling the effect of changing spectrum on current-matching}.
\newblock \emph{\bibinfo{journal}{Opt. Express}} \textbf{\bibinfo{volume}{28}},
  \bibinfo{pages}{7829} (\bibinfo{year}{2020}).
\newblock
  \urlprefix\url{https://www.osapublishing.org/abstract.cfm?URI=oe-28-6-7829}.

\bibitem{geisz_building_2018}
\bibinfo{author}{Geisz, J.~F.} \emph{et~al.}
\newblock \bibinfo{title}{Building a six-junction inverted metamorphic
  concentrator solar cell}.
\newblock \emph{\bibinfo{journal}{{IEEE} Journal of Photovoltaics}}
  \textbf{\bibinfo{volume}{8}}, \bibinfo{pages}{626--632}
  (\bibinfo{year}{2018}).
\newblock \urlprefix\url{https://ieeexplore.ieee.org/document/8231134/}.

\bibitem{parent_2019}
\bibinfo{author}{{Parent}, L.}, \bibinfo{author}{{Riverola}, A.},
  \bibinfo{author}{{Chemisana}, D.}, \bibinfo{author}{{Dollet}, A.} \&
  \bibinfo{author}{{Vossier}, A.}
\newblock \bibinfo{title}{Fine-tuning of multijunction solar cells: An in-depth
  evaluation}.
\newblock \emph{\bibinfo{journal}{IEEE Journal of Photovoltaics}}
  \bibinfo{pages}{1--7} (\bibinfo{year}{2019}).

\bibitem{mcmahon_multijunction_2017}
\bibinfo{author}{McMahon, W.~E.}, \bibinfo{author}{Friedman, D.~J.} \&
  \bibinfo{author}{Geisz, J.~F.}
\newblock \bibinfo{title}{Multijunction solar cell design revisited: disruption
  of current matching by atmospheric absorption bands: {Disruption} of current
  matching by atmospheric absorption bands}.
\newblock \emph{\bibinfo{journal}{Progress in Photovoltaics: Research and
  Applications}}  (\bibinfo{year}{2017}).

\bibitem{hirst_fundamental_2011}
\bibinfo{author}{Hirst, L.~C.} \& \bibinfo{author}{Ekins-Daukes, N.~J.}
\newblock \bibinfo{title}{Fundamental losses in solar cells}.
\newblock \emph{\bibinfo{journal}{Progress in Photovoltaics: Research and
  Applications}} \textbf{\bibinfo{volume}{19}}, \bibinfo{pages}{286--293}
  (\bibinfo{year}{2011}).
\newblock \urlprefix\url{http://doi.wiley.com/10.1002/pip.1024}.

\bibitem{geisz_generalized_2015}
\bibinfo{author}{Geisz, J.~F.} \emph{et~al.}
\newblock \bibinfo{title}{Generalized {Optoelectronic} {Model} of
  {Series}-{Connected} {Multijunction} {Solar} {Cells}}.
\newblock \emph{\bibinfo{journal}{IEEE Journal of Photovoltaics}}
  \textbf{\bibinfo{volume}{5}}, \bibinfo{pages}{1827--1839}
  (\bibinfo{year}{2015}).

\bibitem{chicago_2019}
\bibinfo{author}{Ripalda, J.~M.}, \bibinfo{author}{Buencuerpo, J.} \&
  \bibinfo{author}{García, I.}
\newblock \bibinfo{title}{Dependence of multijunction optimal gaps on spectral
  variability and other environmental and device parameters}.
\newblock In \emph{\bibinfo{booktitle}{IEEE PVSC 46 Proceedings, Chicago}}
  (\bibinfo{year}{2019}).

\bibitem{f_holmgren_pvlib_2018}
\bibinfo{author}{F.~Holmgren, W.}, \bibinfo{author}{W.~Hansen, C.} \&
  \bibinfo{author}{A.~Mikofski, M.}
\newblock \bibinfo{title}{pvlib python: a python package for modeling solar
  energy systems}.
\newblock \emph{\bibinfo{journal}{Journal of Open Source Software}}
  \textbf{\bibinfo{volume}{3}}, \bibinfo{pages}{884} (\bibinfo{year}{2018}).
\newblock \urlprefix\url{http://joss.theoj.org/papers/10.21105/joss.00884}.

\bibitem{de_soto_improvement_2006}
\bibinfo{author}{De~Soto, W.}, \bibinfo{author}{Klein, S.} \&
  \bibinfo{author}{Beckman, W.}
\newblock \bibinfo{title}{Improvement and validation of a model for
  photovoltaic array performance}.
\newblock \emph{\bibinfo{journal}{Solar Energy}} \textbf{\bibinfo{volume}{80}},
  \bibinfo{pages}{78--88} (\bibinfo{year}{2006}).
\newblock
  \urlprefix\url{https://linkinghub.elsevier.com/retrieve/pii/S0038092X05002410}.

\bibitem{ascencio-vasquez_global_2019}
\bibinfo{author}{Ascencio-Vásquez, J.}, \bibinfo{author}{Kaaya, I.},
  \bibinfo{author}{Brecl, K.}, \bibinfo{author}{Weiss, K.-A.} \&
  \bibinfo{author}{Topič, M.}
\newblock \bibinfo{title}{Global climate data processing and mapping of
  degradation mechanisms and degradation rates of {PV} modules}.
\newblock \emph{\bibinfo{journal}{Energies}} \textbf{\bibinfo{volume}{12}},
  \bibinfo{pages}{4749} (\bibinfo{year}{2019}).
\newblock \urlprefix\url{https://www.mdpi.com/1996-1073/12/24/4749}.

\bibitem{vazquez_reliability_2017}
\bibinfo{author}{Vazquez, M.} \emph{et~al.}
\newblock \bibinfo{title}{Reliability of commercial triple junction
  concentrator solar cells under real climatic conditions and its influence on
  electricity cost: {Reliability} of commercial triple junction concentrator
  solar cells}.
\newblock \emph{\bibinfo{journal}{Progress in Photovoltaics: Research and
  Applications}} \textbf{\bibinfo{volume}{25}}, \bibinfo{pages}{905--918}
  (\bibinfo{year}{2017}).

\bibitem{garcia_soiling_2011}
\bibinfo{author}{García, M.}, \bibinfo{author}{Marroyo, L.},
  \bibinfo{author}{Lorenzo, E.} \& \bibinfo{author}{Pérez, M.}
\newblock \bibinfo{title}{Soiling and other optical losses in solar-tracking
  {PV} plants in navarra}.
\newblock \emph{\bibinfo{journal}{Progress in Photovoltaics: Research and
  Applications}} \textbf{\bibinfo{volume}{19}}, \bibinfo{pages}{211--217}
  (\bibinfo{year}{2011}).
\newblock \urlprefix\url{http://doi.wiley.com/10.1002/pip.1004}.

\bibitem{sarver_comprehensive_2013}
\bibinfo{author}{Sarver, T.}, \bibinfo{author}{Al-Qaraghuli, A.} \&
  \bibinfo{author}{Kazmerski, L.~L.}
\newblock \bibinfo{title}{A comprehensive review of the impact of dust on the
  use of solar energy: History, investigations, results, literature, and
  mitigation approaches}.
\newblock \emph{\bibinfo{journal}{Renewable and Sustainable Energy Reviews}}
  \textbf{\bibinfo{volume}{22}}, \bibinfo{pages}{698--733}
  (\bibinfo{year}{2013}).
\newblock
  \urlprefix\url{https://linkinghub.elsevier.com/retrieve/pii/S136403211300021X}.

\bibitem{ilse_techno-economic_2019}
\bibinfo{author}{Ilse, K.} \emph{et~al.}
\newblock \bibinfo{title}{Techno-economic assessment of soiling losses and
  mitigation strategies for solar power generation}.
\newblock \emph{\bibinfo{journal}{Joule}} \textbf{\bibinfo{volume}{3}},
  \bibinfo{pages}{2303--2321} (\bibinfo{year}{2019}).
\newblock
  \urlprefix\url{https://linkinghub.elsevier.com/retrieve/pii/S2542435119304222}.

\bibitem{NREL_2015}
\bibinfo{author}{Sengupta, M.} \emph{et~al.}
\newblock \bibinfo{title}{Best practices handbook for the collection and use of
  solar resource data for solar energy applications}.
\newblock \bibinfo{type}{Tech. Rep.} \bibinfo{number}{NREL/TP-5D00-63112},
  \bibinfo{institution}{NREL} (\bibinfo{year}{2015}).

\bibitem{NREL_Xie_2019}
\bibinfo{author}{Xie, Y.}, \bibinfo{author}{Sengupta, M.},
  \bibinfo{author}{Dooraghi, M.} \& \bibinfo{author}{Habte, A.}
\newblock \bibinfo{title}{Reducing pv performance uncertainty by accurately
  quantifying the pv resource}.
\newblock \bibinfo{type}{Tech. Rep.} \bibinfo{number}{NREL/TP-5D00-73377},
  \bibinfo{institution}{NREL} (\bibinfo{year}{2019}).

\end{thebibliography}

\end{document}